\documentclass[intlimits,twoside,a4paper]{article}
\usepackage[cp1251]{inputenc}

\usepackage{cmpj3}

\usepackage{bm}

\issue{2020}{23}{1}{13603}
\doinumber{10.5488/CMP.23.13603}

\title [The van der Waals idea of pseudo associations and the critical compressibility factor
]{The van der Waals idea of pseudo associations and the critical compressibility 
	factor}
\author[O. Bakai] {O. Bakai\thanks{E-mail: bakai@kipt.kharkov.ua}}
\address{National Science Center ``Kharkiv Institute of Physics \& Technology'', 1 Akademichna St., 61108 Kharkiv, Ukraine}

\date{Received October 2, 2019, in final form October 31, 2019}

\begin{document}

\maketitle

\begin{abstract}
	The dimensionless value of critical compressibility factor in the van der Waals theory 
	of gas-liquid critical point is a universal constant, $Z_\text{c} = 0.375$. Experimentally 
	measured values of this quantity for simple fluids are considerably smaller than 
	the theory prediction. Van der Waals once assumed that this discrepancy can be removed 
	by taking account of the impact of the molecular pseudo-associations on the fluid criticality 
	but he did not complete a proper modification of his theory following up on this 
	idea. The communication is devoted to the filling of this gap.

\end{abstract}

\section{Introduction}

The 
van der Waals equation of state represented in the form of the law of corresponding 
states contains a universal dimensionless ratio, the critical compressibility factor,
\begin{equation}
Z_\text{c} 
=\frac{P_\text{c} v_\text{c} }{T_\text{c} } =0.375,  \label{GrindEQ__1_}
\end{equation}
where $P$, $T$ and $v$ is the pressure, temperature and specific 
volume. Here, index ``c'' denotes quantities evaluated at the critical point. 

 In his famous Nobel prize lecture, discussing the discrepancy between the predicted 
universal value of the critical compressibility factor (\ref{GrindEQ__1_}) and its 
experimentally measured value for CO$_2$  (which is nearly 30\% less), van der Waals 
assumed that it is necessary to take account of the formation of the transient molecular complexes --- 
``pseudo associations'' --- within the dense fluid in order to improve the equation 
of state and the critical compressibility factor. The pseudo-associations were described 
as follows \cite{Waals67}.

 \textit{``Let the number of molecules that have combined into a complex be so large 
that it is possible to speak of a molecule at the centre surrounded by a single layer 
containing almost as many other molecules as is possible simultaneously. Then, for 
the surrounding molecules the attraction directed towards the interior acts only 
to maintain the complex; and this part of its attraction is lost for the surface 
pressure. Only the forces acting outwards from these molecules can contribute to 
the formation of the internal pressure. But of course, for pseudo association as 
for true association the number of formed complexes increases with decreasing temperature 
and volume. At the critical point, so I was compelled to conclude, only a very small 
part of the weight is present as complexes. If pseudo association exists in a substance, 
there are at least two types of molecules, namely simple and complex.'' }

The formulated problem was not properly solved by van der Waals. It is revisited 
in the present communication.

\section{The van der Waals equation of state taking account of the hetero\-phase fluctuations}

In 
fact, the problem formulated by van der Waals was partially solved by Frenkel in 
the theory of heterophase fluctuations (the liquid-like droplets within the gas and 
gaseous bubbles within the liquid) which he formulated considering the pre-transition 
anomalies of the coexisting gas and liquid far below the critical point \cite{Frenkel39}. Identifying 
the liquid-like droplets within the gas as the pseudo-associates, one obtains an 
improved equation of state. However, Frenkel's theory is applicable only in the case 
of weak heterophase fluctuations when the droplets concentration is so small that 
they do not interact with each other (the droplet ideal gas). Frenkel has described 
the impact of the heterophase fluctuations on the thermodynamic quantities far from 
the critical point. 

 Contrary to the van der Waals expectation (see above), near the critical point the 
concentration of droplets is large, and their interaction cannot be ignored (see 
\cite{Bakai2017} and references therein). In this regime, the coexisting nominal gas and nominal 
liquid phases are essentially the states of a heterophase fluid with comparable values 
of the liquid-like and gas-like fractions (l- and g-fractions). 
Elementary constituents of the l- and g-fractions are transient 
droplets and bubbles which have a mesoscopic scale comparable with the range of short-range 
order in the fluid. These entities can be considered as statistically independent species 
of the fluid. They are called l- and g-fluctuons, respectively. 

 The nominal gaseous and liquid phases near the critical point are heterophase mixtures 
of mesoscopic bubbles and droplets. Interactions between heterophase fluctuations 
lead to a modification of the van der Waals equation of state.

 The specific volume of the heterophase fluid, 
\begin{equation}
v=c_\text{g} v_\text{g} +c_\text{l} v_\text{l} \,,\label{GrindEQ__2_}
\end{equation}
is determined by the concentrations of molecules belonging to the bubbles and droplets, $c_\text{g} $ and $c_\text{l} $ respectively. 
Here, $v_\text{g} $ and $v_\text{l} $ is the specific volume in the bubble and droplet, respectively.

 Let us assume that the specific volume (\ref{GrindEQ__2_}) obeys the van der Waals equation 
of state while the l- and g-fluctuons are elementary disturbances.

 The free energy per molecule of a homogeneous one-component simple fluid in the van der 
Waals approximation has a standard form, 
\begin{equation}
 \frac{F_\text{VdW} (v,T)}{N} = -T\ln  \left(\frac{v-b}{\Lambda ^{3} } \right)-\frac{a}{v} \,,   \label{GrindEQ__3_}
\end{equation}
where $\Lambda $ is the thermal de Broglie wavelength, \textit{N} is the number of atoms, $b$ is 
the excluded volume, and $a$ is the virial coefficient.

 To include the contribution of the heterophase fluctuations to the free energy at 
the coexistence curve in a self-consistent manner we have to require that the chemical 
potentials of molecules within the l- and g-fluctuons are equal 
to the chemical potential of the van der Waals fluid. With this condition, the free 
energy of the fluid gets contributions due to the interaction energy, $\varepsilon 
_\text{fl,int} $, and the mixing entropy, $s_\text{fl,mix} $, of the l- and g-fluctuons, 
\begin{equation}
 \frac{F_\text{fl} 
(v,T,\sigma _\text{l} )}{N} =f_\text{fl} (v,T,\sigma _\text{l} )=\frac{N_\text{fl} }{N} \left(\varepsilon 
_\text{fl,{int}} -Ts_\text{fl,{mix}} \right)=\frac{1}{\bar{k}_\text{fl} } \left[g_{2} \sigma _\text{l} \sigma 
_\text{g} +T\left(\sigma _\text{l} \ln \sigma _\text{l} +\sigma _\text{g} \ln \sigma _\text{g} \right)\right],  
\label{GrindEQ__4_}
\end{equation}
where index ``fl'' denotes the quantities describing the fluctuons, $N_\text{fl}$ is 
the number of fluctuons, $\bar{k}_\text{fl}$ is the mean number of atoms per fluctuon, $\sigma 
_\text{l} $ and $\sigma _\text{g} $ are fractions of the l- and g-fluctuons 
($\sigma _\text{l} +\sigma_\text{g} =1$), and parameter $g_{2} > 0$ quantifies the 
fluctuonic interaction. 

 The free energy accounting for the fluctuonic contribution is given by
\begin{align}
 \frac{F(v,T,\sigma _\text{l} )}{N} =-T\ln  \left(\frac{v-b}{\Lambda ^{3} } \right)-\frac{a}{v} 
+\frac{1}{\bar{k}_\text{fl} } \left[g_{2} \sigma _\text{l} \sigma _\text{g} +T\left(\sigma _\text{l} 
\ln \sigma _\text{l} +\sigma _\text{g} \ln \sigma _\text{g} \right)\right].  \label{GrindEQ__5_}
\end{align}
 The number of molecules per l- and g-fluctuon, $k_\text{l}$ and $k_\text{g}$, 
determine the mean number of molecules per fluctuon, $\bar{k}_\text{fl}$, 
\begin{equation}
 \bar{k}_\text{fl} =\sigma _\text{l} k_\text{l} +\sigma_\text{g} k_\text{g}.  \label{GrindEQ__6_}
\end{equation}
The specific volumes of the l- and g-fraction are related to the 
mean specific volume $v$ and $\sigma_\text{l}$ and $\sigma_\text{g}$ as follows:
\begin{equation}
v=\frac{\sigma _\text{l} k_\text{l} v_\text{l} +\sigma _\text{g} k_\text{g} v_\text{g} }{\bar{k}_\text{fl} }.  \label{GrindEQ__7_}
\end{equation}
This equation is equivalent to equation~(\ref{GrindEQ__2_}). Combining equations~(\ref{GrindEQ__5_}) 
and (\ref{GrindEQ__7_}) one has
\begin{align}
\frac{F(v,T,\sigma _\text{l} )}{N} &\equiv f(v,T,\sigma _\text{l} )=  
  -T\ln   \left(\frac{v-b}{\Lambda ^{3} 
} \right)-\frac{a}{v}\nonumber\\
& +\frac{v}{\sigma _\text{l} k_\text{l} v_\text{l} +\sigma _\text{g} k_\text{g} v_\text{g} 
} \left[g_{2} \sigma _\text{l} \sigma _\text{g} +T\left(\sigma _\text{l} \ln \sigma _\text{l} +\sigma 
_\text{g} \ln \sigma _\text{g} \right)\right].  \label{GrindEQ__8_}
\end{align}
As seen from equation~(\ref{GrindEQ__8_}), the heterophase fluctuations affect the pressure,
\begin{align} 
 P&=\frac{\partial f(v,T,\sigma _\text{l} )}{\partial 
v} =\frac{T}{v-b}-\frac{a}{v^{2}}  +\frac{1}{\sigma _\text{l} k_\text{l} v_\text{l} +\sigma _\text{g} 
k_\text{g} v_\text{g} } \left[g_{2} \sigma _\text{l} \sigma _\text{g} +T\left(\sigma _\text{l} \ln \sigma 
_\text{l} +\sigma _\text{g} \ln \sigma _\text{g} \right)\right]
\nonumber\\
&=P_\text{VdW} (v,T)+P_\text{fl} (T,\sigma _\text{l} 
). \label{GrindEQ__9_} 
\end{align} 
Here, $P_\text{VdW} (v,T_\text{l} )$ is the van der Waals pressure and
\begin{equation} \label{GrindEQ__10_} P_\text{fl} (T,\sigma _\text{l} )=\frac{1}{\sigma _\text{l} 
k_\text{l} v_\text{l} +\sigma _\text{g} k_\text{g} v_\text{g} } \left[g_{2} \sigma _\text{l} \sigma _\text{g} +T\left(
\sigma _\text{l} \ln \sigma _\text{l} +\sigma _\text{g} \ln \sigma _\text{g} \right)\right] 
\end{equation} 
is the fluctuonic pressure. 

 The equilibrium value of $\sigma _\text{l} $ is found by minimizing $f\left(v,T,\sigma 
_\text{l} \right)$ with respect to this parameter.  One should find solutions of the equation 
\begin{equation}
 \frac{
\partial f\left(v,T,\sigma _\text{l} \right)}{\partial \sigma _\text{l} } =0,
  \label{GrindEQ__11_}
\end{equation}
that simultaneously satisfy the condition
\begin{equation}
 \frac{\partial ^{2} f\left(v,T,\sigma _\text{l} \right)}{\partial \sigma _\text{l}^{2} } >0.  \label{GrindEQ__12_}
\end{equation}

To this end, the Ising-type effective Hamiltonian, $g_{2} \sigma _\text{l} \sigma _\text{g} 
+T\left(\sigma _\text{l} \ln \sigma _\text{l} +\sigma _\text{g} \ln \sigma _\text{g} \right)$, should 
be minimized. As known, it possesses the critical (bifurcation) temperature $T_\text{c} 
=g_{2} /2$ at which $\sigma _\text{l} (T_\text{c} )=\sigma _\text{g} (T_\text{c} )=1/2$. For consistency 
with the van der Waals theory we have to set 
\begin{equation} 
\label{GrindEQ__13_} 
g_{2} =2T_{\text{c,VdW}} =\frac{16}{27} \frac{a}{b}  
\end{equation} 
and, as it follows from equations~(\ref{GrindEQ__6_}), (\ref{GrindEQ__7_}), 
\begin{equation} 
 \bar{k}_\text{fl,c} =\frac{k_\text{l,{c}} +k_\text{g,{c}} }{2} \,,\quad v_\text{c} =\left({\rm 1}-\varsigma 
\varsigma _{v} \right)\frac{v_\text{l,{c}} +v_\text{g,{c}} }{2} .  
\label{GrindEQ__14_}
\end{equation} 
Here, 
\begin{equation} 
\varsigma =\frac{k_\text{l} -k_\text{g} }{k_\text{l} +k_\text{g} }\,, \quad \varsigma _{v} =\frac{v_\text{g} 
-v_\text{l} }{v_\text{l} +v_\text{g} }.  
\label{GrindEQ__15_}
\end{equation}

Let us note that $f_\text{fl} (v_\text{c},T_\text{c} )<0$. It means that the homogeneous fluid 
is unstable. It transforms into a stable heterophase fluid consisting of  the l- 
and g-fluctuons. At the critical point, the fluctuonic contribution to pressure 
is equal to
\begin{equation}
 P_\text{fl,c} =\frac{2}{k_\text{l,c} v_\text{l,c} +k_\text{g,c} v_\text{g} } \left(\frac{1}{4} g_{2} -T_\text{c} 
\ln 2\right)=-\frac{T_\text{c} \left(\ln 2-0.5\right)}{\left(1-\varsigma _\text{c} \varsigma 
_{v,\text{c}} \right)} \frac{2}{v_\text{l,c} +v_\text{g,c} } =  -0.193\frac{T_\text{c} 
}{\bar{k}_\text{fl,c} v_\text{c} }.
 \label{GrindEQ__16_}
\end{equation}
Thus, the total critical pressure is equal to
\begin{equation}
 P_\text{c} =\frac{a}{27b^{2} } -0.193\frac{T_\text{c} }{\bar{k}_\text{fl,c} v_\text{c} }\,,
 \label{GrindEQ__17_}
\end{equation}
and the fluctuonic pressure impacts the critical compressibility factor
\begin{equation}
Z_\text{c} =\frac{P_\text{c} v_\text{c} }{T_\text{c} } =Z_\text{c,VdW} -\frac{0.193}{\bar{k}_\text{fl,c} } =0.375-
\frac{0.193}{\bar{k}_\text{fl,c} }.  \label{GrindEQ__18_}
\end{equation}

\section{Recovering $\bar{k}_\text{fl,c}$ from experimental data}

It is seen that the fluctuonic correction in equation~(\ref{GrindEQ__18_}) depends on 
just one mesoscopic parameter $\bar{k}_\text{fl,c}$ and reduces the critical compressibility 
factor.  Assuming that the difference between $Z_\text{c,VdW}$ and the measured value $Z_{\text{c},\exp}$, 
is due to the heterophase fluctuations, one can estimate $\bar{k}_\text{fl,c} $ using equation~
(\ref{GrindEQ__18_})
\begin{equation}
 \bar{k}_\text{fl,c} =\frac{0.193}{Z_\text{c,VdW} -Z_{\text{c},\exp } }.  \label{GrindEQ__19_}
\end{equation}
Results for some fluids with London dispersion intermolecular forces are represented in 
table~\ref{tab_bakai1}.
\begin{table}[h]
	\caption{ Values of the parameter $\bar{k}_\text{fl,c} $ for fluids with dispersion intermolecular forces.}
	 \centering
	 \vspace{2ex}
	 \begin{tabular}{|c|c|c|c|} 
	 	\hline \hline
	Fluid &  $T_\text{c}$, K & $Z_{\text{c},\exp}= \frac{P_\text{c}v_\text{c}}{T_\text{c}}$ & $\bar{k}_\text{fl,c} $ \\ \hline \hline
	C$_2$H$_4$ & 282 & 0.270 & 1.83 \\ \hline 
	CO$_2$ & 304.19 & 0.275 & 1.93 \\ \hline 
	Xe & 289.8 & 0.288 & 2.2 \\ \hline 
	Kr & 209.48 & 0.288 & 2.2 \\ \hline 
	CH$_4$ & 190.8 & 0.290 & 2.27 \\ \hline 
	N$_2$ & 126.2 & 0.291 & 2.27 \\ \hline 
	Ar & 150.8 & 0.291 & 2.27 \\ \hline 
	O$_2$ & 154.6 & 0.292 & 2.32 \\ \hline 
	Ne & 44.5 & 0.298 & 2.47 \\ \hline 
	H$_2$ & 33.2 & 0.304 & 2.7 \\ \hline 
	He$_4$ & 5.20 & 0.308 & 2.88 \\ \hline 
	He$_3$ & 5.19 & 0.320 & 3.5 \\ \hline \hline
	\end{tabular}
	\label{tab_bakai1}
\end{table}

 The obtained values of $\bar{k}_\text{fl,c} $ can be used for making specification of the parameters 
of mesoscopic models of fluid \cite{Bakai2017}. Let us note that the Lee-Yang lattice-gas model 
(LY-model) \cite{Lee52} is a partial case of the lattice-fluctuon model (LFM) \cite{Bakai2017} and that 
the last one reproduces the results of the LY-model at $ k_\text{l} =1$, $k_\text{g} =0$ and $\varsigma =\varsigma_{v}=1$. In the LY-model $\bar{k}_\text{fl,c} 
=k_\text{l} /2=1/2$.  In the LFM $\varsigma ,\varsigma _{v} <1$ and $k_\text{l} <2\bar{k}_\text{fl,c} $. Therefore,  $k_\text{l} <3.66 $ for ethylene and $k_\text{l} < 7$ for He$^{3}$.  

 A growth of $\bar{k}_\text{fl,c} $ for the quantum liquids H$_2$,  He$^4$ and He$^3$ correlates 
with an increase of the de Boer parameter.  Tabulated values of $B$ can be found, 
e.g., in \cite{Apfelbaum09}. For H$_2$,  He$^4$ and He$^3$, the parameter $B$ is equal to $0.78$, $0.99$ and $1.76$, 
respectively. Correlation between $\bar{k}_\text{fl,c} $ and $B$ can be interpreted 
as follows. Since the l-fluctuon radius is nearly equal to the correlation 
radius of the direct correlation function, it increases with an increase of the 
parameter $B$ that characterizes the molecule delocalization. In the Kac model~\cite{Kac63}, the van der Waals equation is obtained for the system of hard spheres with the 
range of the attractive force tending to infinity while its strength becomes proportionally 
weaker. The parameter $B$ increases with the growth of the attractive force range and 
tends to infinity. Therefore, the compressibility parameter  expectedly approaches $Z_\text{c,VdW}$  and the 
accuracy of the van der Waals equation improves with an increase of the parameter 
$B$.

\section{Discussion and summary}

The success of the law of the corresponding states urged van der Waals to look for a 
general physical reason of the critical compressibility factor non-universality. This 
problem is a subject of permanent interest. A number of recent papers are devoted 
to this problem (see \cite{Apfelbaum09,Apfelbaum13,Kulinskii14} and references quoted). Considerations reported here 
show that  the van der Waals conjecture on the role of pseudo associates is reasonable. 
Moreover,  it allows one to get an estimate of the number of molecule within the l-fluctuon 
which is useful while applying the  mesoscopic models of gas-liquid transformations 
\cite{Bakai2017}.

To conclude, reformulation of the van der Waals equation of state taking account of 
the heterophase fluctuations in the vicinity of the critical point allows one to 
associate the critical compressibility factor with the mesoscopic parameters of the l- 
and g-fluctuons. The formulated equation revives the idea of the pseudo-associates impact on the critical compressibility factor.

\section*{Acknowledgement} 

 The author is deeply grateful to Professor J.M.H. Levelt Sengers for providing the 
text of her excellent historical overview of the problem of fluid heterogeneities 
near the critical point \cite{Levelt79}.

\ukrainianpart

\title{Ідея псевдоасоціації Ван дер Ваальса  і  коефіцієнт критичної стисливості}
\author{O. Бакай}
\address{Національний науковий центр  ``Харківський фізико-технічний інститут'', \\вул. Академічна, 1, 61108 Харків, Україна}

\makeukrtitle

\begin{abstract}
	Безрозмірний коефіцієнт критичної стисливості в теорії Ван дер Ваальса критичної точки газ-рідина є універсальною константою, $Z_\text{c} = 0.375$. Експериментально виміряне значення цієї величини для простих плинів є значно меншим, ніж теоретичне передбачення. Ван дер Ваальс зробив передбачення, що ця розбіжність може бути усунута, якщо врахувати вплив молекулярної псевдоасоціації на критичність плину. Проте,  він не завершив модифікації своєї теорії відповідно до цієї ідеї. Ця робота присвячена заповненню цієї прогалини.

	\keywords  Ван дер Ваальс, коефіцієнт критичної стисливості, гетерофазні флуктуації 
\end{abstract}

\end{document}